\documentclass[twocolumn,prx,aps,longbibliography]{revtex4-1}

\usepackage{graphicx}
\usepackage{float}
\usepackage{amssymb,amsthm,amsfonts,amstext,amsmath,amssymb}
\usepackage{url}
\def\be{\begin{equation}}
\def\ee{\end{equation}}
\def\bea{\begin{eqnarray}}
\def\eea{\end{eqnarray}}
\def\bma{\begin{mathletters}}
\def\ema{\end{mathletters}}

\def\q0{\underline{0}}

\def\H{{\cal H}}

\def\C{{\mathbb C}}
\def\id{{\mathbb I}}

\def\H{{\cal H}}

\def\R{\mathbb{R}}
\def\N{\mathbb{N}}

\def\tr{\mbox{tr}}
\def\one{\leavevmode\hbox{\small1\normalsize\kern-.33em1}}

\def\bra#1{\langle#1|} \def\ket#1{|#1\rangle}
\def\braket#1#2{\langle#1|#2\rangle}

\def\proj#1{\ket{#1}\!\bra{#1}}

\def\id{{\mathbb I}}

\begin{document}

\title{Resetting uncontrolled quantum systems}
\author{Miguel Navascu\'es}
\affiliation{Institute for Quantum Optics and Quantum Information (IQOQI) Vienna, Austrian Academy of Sciences, Boltzmanngasse 3, 1090 Vienna, Austria}

\begin{abstract}
We consider a scenario where we wish to bring a closed system of known Hilbert space dimension $d_S$ (the target), subject to an unknown Hamiltonian evolution, back to its quantum state at a past time $t_0$. The target is out of our control: this means that we ignore both its free Hamiltonian and how the system interacts with other quantum systems we may use to influence it. Under these conditions, we prove that there exist protocols within the framework of non-relativistic quantum physics which reset the target system to its exact quantum state at $t_0$. Each ``resetting protocol'' is successful with non-zero probability for all possible free Hamiltonians and interaction unitaries, save a subset of zero measure. When the  target is a qubit and the interaction is sampled from the Haar measure, the simplest resetting circuits have a significant average probability of success and their implementation is within reach of current quantum technologies. Finally, we find that, in case the resetting protocol fails, it is possible to run a further protocol that, if successful, undoes both the natural evolution of the target and the effects of the failed protocol over the latter. By chaining in this fashion several such protocols, one can substantially increase the overall probability of a successful resetting.
\end{abstract}

\maketitle

In Newtonian Physics, as well as in non-relativistic quantum theory, time is regarded as a real external parameter that is not subject to dynamics but describes the evolution of the whole universe in Newtonian absolute space. This leaves out the possibility to influence or manipulate it in any way. 

This notion of time, however, does not correspond to the entity that we measure in the lab when we speak of, e.g., the time between a particle's production and its subsequent detection. In real life, to measure time, we use \emph{clocks}, i.e., physical devices whose state describes a trajectory $\{\psi(s):s\in\R\}$ in state space. When we say that an event happened at time $s$, what we actually mean is that the state of our clock was $\psi(s)$ when we recorded the event \footnote{In quantum theory, we should assume that the clock is macroscopic, so that a precise reading of $s$ can be made without perturbing $\psi(s)$ too much. Otherwise, we should resort to an imprecise estimation of $s$ through a weak measurement.}. 

Syncronizing two different clocks, with state trajectories $\{\psi_1(s_1):s_1\}$, $\{\psi_2(s_2):s_2\}$ amounts to party $1$ ($2$) being able to predict the state $\psi_2$ ($\psi_1$) of clock $2$ ($1$), given the state of its own clock $\psi_1$ ($\psi_2$). Mathematically, this involves identifying a parametrization $s_1(s),s_2(s)$ such that, for all values of $s$, the simultaneous state of both clocks is $\psi_1(s_1(s)),\psi_2(s_2(s))$, respectively. If both parties now parametrize time by $s$, then they will always agree on the time coordinate of each event. This other conception of time, as a relation between the physical states of different systems, is at the heart of our understanding of time in general relativity and in the relational approach of dynamics in quantum gravity \cite{rovelli}. 

It also opens the door to manipulating the local time within a physical system or ``moving it through time''. Indeed, consider a physical system $S$ that, left to its own, would follow a trajectory in state space $\{\psi(t):t\}$, where $t$ denotes the agreed parametrization of time between us and $S$. Now, imagine that we interacted with $S$ within the time interval $t\in [0, \tau]$ in such a way that, at the end of our interaction, the state of the system were not $\psi(\tau)$, but $\psi(\tau')$, with $\tau'\not=\tau$. If, from this point on, $S$ kept evolving as expected, then there would be a mismatch $\tau-\tau'$ between the time measured in our lab and at system $S$. From our point of view, there would have been a disruption in the normal flow or progress of time in system $S$. Following Merriam-Webster \cite{merriam}, we will call such a disruption a \emph{time-warp}.

There exist a number of proposals to carry out this effect. For instance, special relativity teaches us that we can decrease the flow of time within a physical system just by accelerating and decelerating it with respect to us. This would allow us to carry a time-warp experience with $0<\tau'<\tau$. Such a form of time-warp is very limited, though: it doesn't allow us to increase the flow of time within the system ($\tau'>\tau$), or to reverse its direction ($\tau'<0$). Further schemes based on general relativity have been proposed to achieve these other effects \cite{goedel,thorne}. In this regard, it is worth mentioning the interesting -and inexplicably unnoticed--- work of \cite{sandu}, where the authors show how to use linear superpositions of space-time metrics to accelerate, slow or reverse the unitary time evolution of a closed quantum system.

At this point it is worth remarking there is nothing mysterious about time-warp in itself. Consider, e.g., a gas of classical particles. We could study its dynamics via a Maxwell demon, and, once we had an accurate Hamiltonian description of the particles, ask the demon to return all of them to the position they must have occupied a year ago. That would qualify as time-warp, according to our definition. However, this time-warp scheme requires an absolute control of the considered system. The merit of the proposals above is, precisely, that they effect time-warp within a system over which we hardly have any control. Coming back to special relativity, one does not need to act precisely on each internal degree of freedom of a system in order to dilate its proper time: it is enough to give the system a push.

Unfortunately, the time-warp schemes listed so far turn out to be highly impractical, when not impossible, to realize. We do observe time dilation in particle accelerators and orbiting satellites, but large effects can just be measured at a sub-atomic level (and at a vast energy cost). Time-warping schemes based on relativistic time-travel seem to require the violation of a number of basic physical principles \cite{visser, hawking}. Technological challenges aside, the scheme proposed in \cite{sandu} to perform time-translations in quantum systems has an astronomically small probability of success. In the words of one of its co-authors: ``it has the same chances of succeeding as I have of delocalizing and relocalizing somewhere else'' \footnote{L. Vaidman, private communication.}.

In this paper, we investigate the feasibility of time-warp from the point of view of non-relativistic quantum theory. Surprisingly, we find that it is possible to engineer particle beams with the property to project the systems they interact with to a past quantum state. More specifically: the target system, whose free evolution is governed by an \emph{unknown} time-independent Hamiltonian, is made to interact sequentially with a number of quantum probes. By manipulating these probes before and after their \emph{also unknown} interaction with the target, we induce a heralded probabilistic transformation on the latter, that, if successful, will bring its quantum state back to the one it had at an arbitrarily long time before we started the ``resetting protocol''. In the language above, the physical realization of any such protocol can be interpreted as a time-warp experience with $\tau'<0$.

Contrarily to the time-translation scheme proposed in \cite{sandu}, our resetting protocols exhibit a significant probability of success. In addition, should the resetting protocol fail, it is possible to carry a further protocol to revert both the past unitary evolution of the system and the action of the first protocol over the target. By iterating this procedure a few times, one can considerably increase the probability of a successful reset. Finally, the simplest protocols just require control over three qubits and hence can be implemented with current quantum technologies.

\vspace{10pt}
\noindent\textbf{The scenario}

Think of a quantum system $S$ (our target) of dimension $d_S$, undergoing an evolution determined by an (unknown) time-independent Hamiltonian $H_0$. Acting on $S$ from time $t=T>0$, we wish to reset the current state of the target, $\ket{\psi(T)}=e^{-iH_0T}\ket{\psi(0)}$, to its past value $\ket{\psi(0)}$.

\begin{figure}
  \centering
  \includegraphics[width=9 cm]{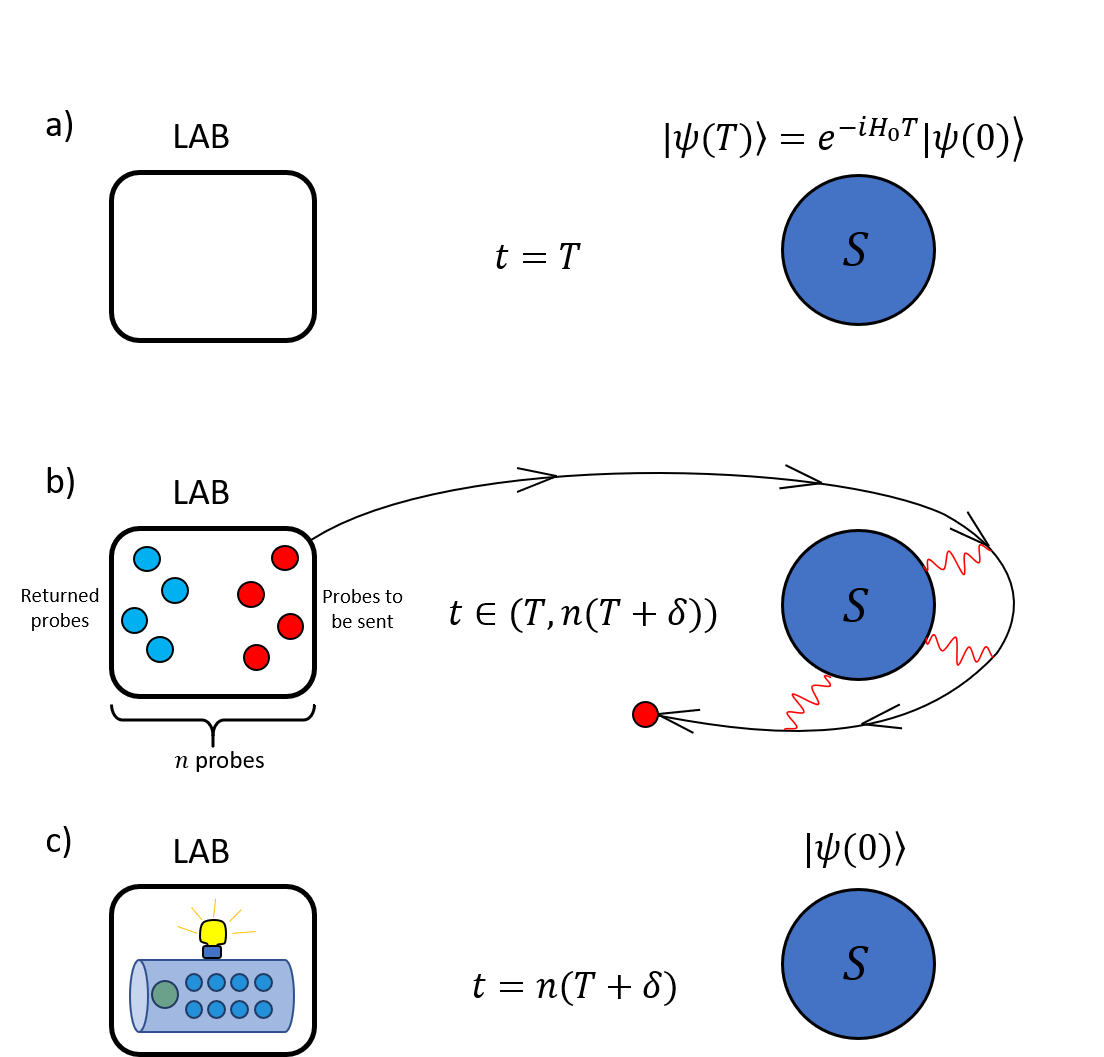}
  \caption{\textbf{A quantum resetting protocol.} a) At time $t=T$, the state of the target system $S$ is the result of evolving the state $\ket{\psi(0)}$ for time $T$ via the unknown Hamiltonian $H_0$. b) At times $t=T,2T+\delta,3T+2\delta...$, the lab sets a probe in a trajectory close to system $S$, that interacts with it in an unknown (unitary) way for a time $\delta$. For times $t\in [T+\delta,2T+\delta], [2T+2\delta,3T+2\delta],...$, the evolution is once more governed by the free Hamiltonian $H_0$. c) Conditioned on some heralded probabilistic operation on the returned probes (in blue) and possibly an extra ancillary system (in green), the state of system $S$ at time $n(T+\delta)$ is again $\ket{\psi(0)}$.}
  \label{reset}
\end{figure}

There is an additional complication: the target system is \emph{uncontrolled}. This means that we ignore how the target evolves by itself (i.e., we ignore $H_0$) and how it jointly evolves with other quantum systems we may use to influence it. We can picture this scenario by imagining that the target is outside our perfectly controlled quantum lab and our only means of interacting with it is by setting a quantum probe $P$ in an orbit close to $S$ and then back to the lab, for a total amount of time $\delta$, see Figure \ref{reset}. When we do so, the joint state $\rho_{SP}$ of both target and probe will have evolved according to an \emph{unknown} joint unitary $W_{S P}$, the result of integrating the evolution equation $i\frac{d\rho_{SP}}{d s}=[H_0+H_{SP}(\bar{r}(s)),\rho_{SP}]$ from $s=0$ to $s=\delta$. Here $H_{SP}(\bar{r})$ denotes the (unknown) interaction between target and probe, that depends implicitly on time through the relative position $\bar{r}$ between $S$ and $P$.

To return the target system to its original state, we will carry out the \emph{quantum resetting protocol} sketched in Figure \ref{reset}: at time $T$, we will send a first probe to system $S$. Upon its return to the lab, at time $T+\delta$, we will prepare a second probe, that we will keep in the lab for $T$ time units and then send in an identical orbit around $S$ at time $2T+\delta$. After this second probe arrives, we again prepare a third probe, wait for time $T$ and then send it around $S$. We iterate this procedure until the $n^{th}$ probe arrives at the lab at time $t_f\equiv n(T+\delta)$. By conducting an operation over the returned probes and possibly on some extra ancillary system, we wish to project system $S$ to its exact quantum state at time $t=0$. 

The first probe, in principle, can be prepared at the beginning of the protocol (i.e., at time $T$), but for theoretical convenience and w.l.o.g. we will \emph{pretend} that the probe already existed in the lab at time $t=0$ and did not evolve between times $[0,T]$. That way, for $k=1,...,n$ the joint evolution of the target and the $k^{th}$ probe from time $(k-1)(T+\delta)$ to $k(T+\delta)]$ can be modeled with the same bipartite unitary operator $U\equiv W_{SP}(V_S\otimes \id_P)$, with $V\equiv e^{-iH_0T}$ (namely, a solo evolution of system $S$ for time $T$ followed by an interaction between $S$ and $P$ lasting $\delta$ seconds).

\begin{figure}
  \centering
  \includegraphics[width=7 cm]{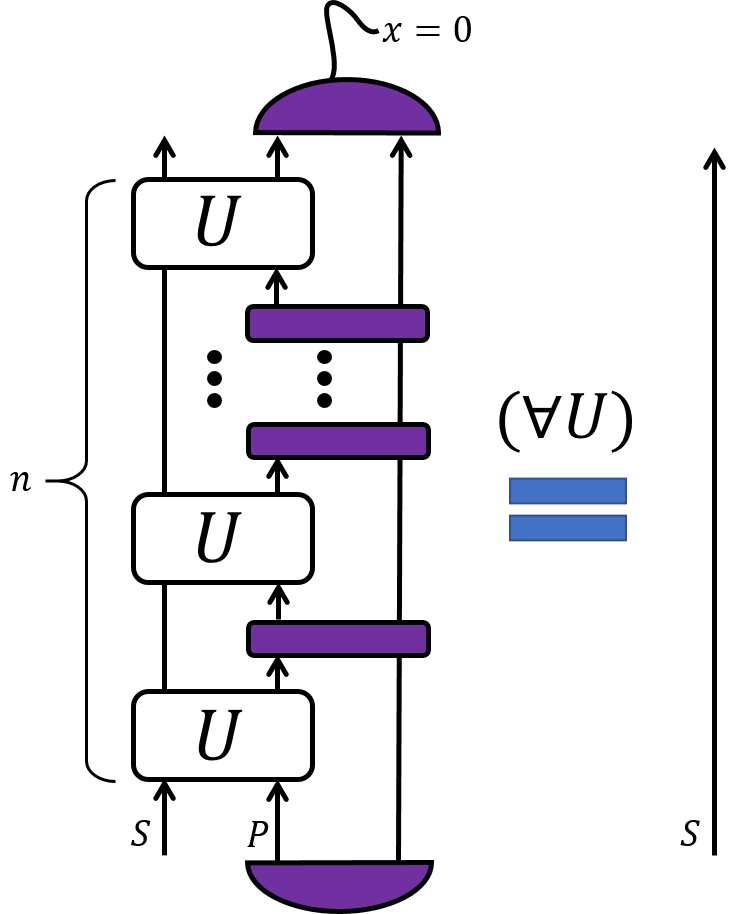}
  \caption{\textbf{Process diagram for a quantum resetting protocol.} Lines denote quantum systems; cups, quantum state preparations; caps, measurements; and rectangles, unitary transformations. Conditioned on a positive result $x=0$, the action of the purple tester is to propagate the original state of the target to the future.}
  \label{sketch}
\end{figure}

This allows us to reformulate the resetting scenario as a quantum network of the form depicted in white in Figure \ref{sketch}, whereby the target and each probe undergo the joint evolution $U$, and we can choose how to prepare each probe before each unitary interaction and what we do with it after it returns to the lab. Our hope is that there exists a quantum circuit that produces inputs for the network and processes its outputs in such a way that the final effect on system $S$ is simply to undo the effect of the interaction $U$, see the purple device in Figure \ref{sketch}.

Clearly, this hypothetical circuit cannot be deterministic. Suppose that the unknown unitary $U_{SP}$ happened to be of the form $U_{SP}=V_S\otimes \tilde{V}_P$, i.e., such that systems $S$ and $P$ do not interact at all. Then, no matter what operations we apply to our probes, the state of system $S$ at time $t_f$ will be $\ket{\psi(t_f)}=V_S^n\ket{\psi(0)}$. If $U_{SP}$ is just \emph{close} to being a tensor product, then, by continuity, the probability of success of our undoing operation cannot be $1$.

These considerations imply that, at best, our scheme will have the effect of leaving the target in state $\ket{\psi(t=0)}$ with a probability $p(U)$ that depends on the particular joint evolution $U$ between system and probe. In particular, $p(U)=0$ when $U$ is of tensor product form. In other words: this circuit, it it exists, will involve conducting a heralding measurement, that we can postpone till the end of the protocol. The outcome $x$ of this binary measurement, say, $0$ or $1$, will tell us whether the state of $S$ remains the same it was at time $t=0$ ($x=0$) or, on the contrary, it has experienced an uncontrolled evolution ($x=1$). In the quantum network slang, such a circuit is termed a \emph{quantum tester} \cite{tester}, \cite{tester2}. 

We remark once more that we demand this tester to be sound under \emph{all} unitaries $U$. That is, whenever the final outcome $x$ of the tester is $0$, we must guarantee that the state of system $S$ is \emph{exactly} $\ket{\psi(0)}$ independently of the particular unitary interaction $U$ that guided the evolution. Due to this soundness requirement, one can prove that quantum resetting is impossible if we allow $U$ to be an arbitrary trace-preserving completely positive map, see Appendix \ref{characterization}. 


At this point, it is interesting to draw a comparison between this scenario and that of \emph{refocusing}, see, e.g., \cite{spinEcho, dynamicalDecoupling}. Given a quantum system $S$ in state $\rho(0)$, subject to a partially unknown interaction with an environment $E$, the purpose of refocusing is to make sure that, at some fixed time $\Delta$, the state of system $S$ will still be approximately $\rho(0)$. This is achieved by applying over $S$ a fast sequence of unitary pulses for time $t\in [0,\Delta]$: if the interaction with the environment is weak enough, the net effect of the pulses will be to freeze the evolution of system $S$. Contrarily to resetting, standard refocusing techniques do not allow us to project the state $\rho(0)$ of $S$ to $\rho(-T)$ for any finite amount of time $T>0$. 

The related primitive \emph{universal unitary refocusing} is another story \cite{refocusing}. Here, as in resetting, it is assumed that the target system has evolved under the action of an unknown time-independent Hamiltonian $H_0$ for a finite time $T$, i.e., via the unitary operator $V=e^{-iH_0T}$. A universal unitary refocusing protocol consists in alternating a sequence of unitary gates $U_1,...,U_n$ with the natural evolution of the target such that $U_nVU_{n-1}...U_2VU_1 \ket{\psi(T)}\approx \ket{\psi(0)}$ with very high probability. The main difference between universal unitary refocusing (and also standard refocusing) and resetting protocols is that in the former the target system is controlled, i.e., we can act on it with any quantum operation we wish (such as $U_1,...,U_n$). In those conditions, it is not difficult to reverse or accelerate the evolution of the system: one could in principle move the state of the system to a quantum memory, find the unknown unitary via channel tomography, apply it or its inverse several times on the modified state of the system and finally move back the resulting state to system $S$. The merit of \cite{refocusing} is to show that one does not need ancillas to accomplish this. While technically interesting, this feature is irrelevant from the point of view of time-warp.

The next question is whether quantum resetting protocols actually exist. We will next show an explicit construction for the simplest non-trivial scenario: $n=4, d_S=d_P=2$.

\vspace{10pt}
\noindent\textbf{A simple quantum resetting protocol}

\begin{figure}
  \centering
  \includegraphics[width=4 cm]{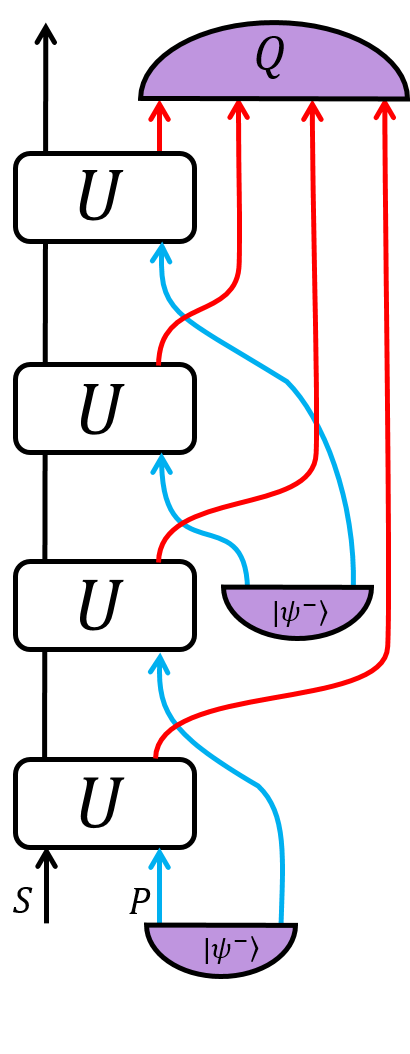}
  \caption{\textbf{A resetting protocol for $n=4, d_S=d_P=2$.} We input two copies of the singlet state and then project the four output qubits onto the quasy-symmetric space $Q$. If the projection succeeds, the state of the target does not vary.}
  \label{proto}
\end{figure}

Suppose then that both systems $S$ and $P$ are qubits, and that we wish our device to return $S$ to its original state after making it interact sequentially with $n=4$ probes (when the target is a qubit, it can be proven that no resetting protocol exists for $n=3$). A possible prototype to do the job is the one depicted in Figure \ref{proto}. We call this protocol ${\cal W}_4$.

Let $\{\ket{0},\ket{1}\}$ be an orthonormal basis for the two-dimensional Hilbert spaces where $S$ and $P$ live. As shown in Figure \ref{proto}, it suffices to prepare our four probes $P_1,P_2,P_3,P_4$ in the quantum state $\ket{\psi^{-}}_{12}\ket{\psi^{-}}_{34}$, where $\ket{\psi^{-}}$ is the singlet state $\ket{\psi^{-}}=\frac{1}{\sqrt{2}}(\ket{0}\ket{1}-\ket{1}\ket{0})$. After their return, probes $1,2,3,4$ are post-selected to the \emph{quasi-symmetric space} $Q$ generated by the vectors

\begin{eqnarray}
&\ket{m_1}=\ket{0,0,0,0},\nonumber\\
&\ket{m_2}=\frac{1}{2}(\ket{1,0,0,0}+\ket{0,1,0,0}+\ket{0,0,1,0}+\ket{0,0,0,1}),\nonumber\\
&\ket{m_3}=\frac{1}{2}(\ket{1,0,1,0}+\ket{0,1,0,1}+\ket{1,0,0,1}+\ket{0,1,1,0}),\nonumber\\
&\ket{m_4}=\frac{1}{\sqrt{2}}(\ket{0,0,1,1}+\ket{1,1,0,0}),\nonumber\\
&\ket{m_5}=\frac{1}{2}(\ket{1,1,1,0}+\ket{0,1,1,1}+\ket{1,0,1,1}+\ket{1,1,0,1}),\nonumber\\
&\ket{m_6}=\ket{1,1,1,1},
\label{quasi_sym}
\end{eqnarray}
\noindent where the state $\ket{a_1,a_2,a_3,a_4}$ must be understood as $\ket{a_1}_{1}\ket{a_2}_{2}\ket{a_3}_{3}\ket{a_4}_{4}$. To get a grasp on the structure of $Q$, note that the symmetric subspace of two qubits is spanned by the vectors $\ket{\overline{0}}\equiv\ket{00},\ket{\overline{1}}\equiv\frac{1}{\sqrt{2}}(\ket{0,1}+\ket{1,0}),\ket{\overline{2}}\equiv\ket{11}$. $Q$ corresponds to the symmetric space of two copies of two symmetric qubits, i.e., it is spanned by the vectors $\{\ket{\overline{k}}\ket{\overline{l}}+\ket{\overline{l}}\ket{\overline{k}}:k,l=0,1,2\}$.

The result of post-selecting the returned probes on any of the states $\{\ket{m_i}\}_{i=1}^6$ is to multiply the wave function of system $S$ by a homogeneous polynomial of degree $n=4$ on the $2\times 2$ complex matrices $\{U_{j,k}\}_{j,k=0,1}$, where 

\be
U_{k,l}\equiv\sum_{i,j=0,1}\bra{j}_S\bra{l}_PU\ket{i}_S\ket{k}_P\ket{j}\bra{i}.
\label{partU}
\ee

\noindent For example, if we post-select the output probe qubits to the state $\ket{m_1}$, then the (non-normalized) state of system $S$ at time $t_f$ will be

\be
\ket{\psi(t_f)}=\frac{1}{2}[U_{0,0},U_{0,1}]^2\ket{\psi(0)}.
\ee

Now, note that $[U_{0,0},U_{0,1}]$ can be expressed as a linear combination of Pauli matrices $X,Y,Z$, i.e., $[U_{0,0},U_{0,1}]=c_xX+c_yY+c_zZ$ (the contribution of $\id_2$ is missing because $\tr([U_{0,0},U_{0,1}])=0$). If we now square this operator we arrive at $[U_{0,0},U_{0,1}]^2=(c^2_x+c^2_y+c^2_z)\id_2$. The matrix polynomial $\frac{1}{2}[U_{0,0},U_{0,1}]^2$ is a \emph{central polynomial for dimension $2$} \cite{polyId}, i.e., a polynomial that is proportional to the identity when evaluated with $2\times 2$ matrices $U_{0,0},U_{0,1}$.There exist no central polynomials for dimension $d_S=2$ with degree smaller than $4$, not even when we restrict the matrices $U_{k,l}$ to be pieces of two-qubit unitaries, and hence a resetting protocol for $n<4$ is impossible.

It can be verified that post-selection with the remaining elements of the basis (\ref{quasi_sym}) also leads to central polynomials acting on $S$\footnote{In fact, all of them follow from the expression $\{[A,B],[C,D]\}_{+}\propto \id_2$. Here $\{\bullet,\bullet\}_{+}$ denotes the anti-commutator, i.e., $\{\alpha,\beta\}_{+}= \alpha\beta+\beta\alpha$.}. As long as we post-select on one of these vectors, we thus have that system $S$ will just acquire a global phase, i.e., it won't change. We note that the state of system $S$ \emph{does} change, though, during the course of the protocol (i.e., it is not frozen). The projection on the space $Q$ spanned by vectors (\ref{quasi_sym}) acts as a sort of \emph{quantum {\bf Ctrl+z}}, undoing such an evolution.

The probability of success of protocol ${\cal W}_4$ just depends on $U$, and not on the initial state of the target. This is a common feature of all quantum resetting protocols. $p(x=0|U,{\cal W}_4)$ varies wildly with $U$: it ranges from zero (for product unitaries) to one (e.g., for $U=\frac{X\otimes Z+iY\otimes X}{\sqrt{2}}$). If $U$ is taken uniformly according to the Haar measure, we find that the average probability of success $\int dU p(x=0|U,{\cal W}_4)$ is approximately $0.2170$. Note that, if the interaction $W_{SP}$ between probe and target is sampled from the Haar measure, so will be $U=W_{SP}(V_S\otimes\id_P)$, independently of $V_S$. Hence the average probability of success will not depend on neither $H_0$ nor $T$.

The ideas behind the above construction can be generalized to show that there exist quantum resetting protocols for target systems of arbitrarily high dimension $d_S$ involving at most $O(d_S^3)$ qubit probes, see Appendix \ref{existence}. There the reader can also find a semidefinite programming \cite{sdp} characterization of the set of resetting protocols for fixed $n, d_S,d_P$. This characterization relies on the theory of quantum testers \cite{tester}, \cite{tester2} and on a variant of the method proposed in \cite{MPS} to compute the support of cut-and-glue operators for homogeneous matrix product states. 

In Appendix \ref{heuristics} we also carry out a comparison between the performance of all quantum resetting protocols with $n=4$ probes and two extremal protocols with $n=8$ in realistic physical scenarios. The results are paradoxical: in some situations, as one would expect, the average probability of success decreases as we try to make the system leap to a more distant past. In some others, the average success probability \emph{grows} with $T$. In addition, we observe that no single extremal protocol outperforms all the others in all situations. The decision to use one protocol or another will depend on our prior knowledge on the target and its interaction with the probes.

\vspace{10pt}
\noindent\textbf{Undoing failure}

Being probabilistic, it is expected that sometimes a resetting protocol yield a negative outcome. Think of protocol ${\cal W}_4$: at time $t_f$, we conduct a projection onto the quasi-symmetric space $Q$, see eq. (\ref{quasi_sym}). Suppose that the said projection fails, but we conducted it in a non-demolition way. Let $\{\ket{\tilde{m}_i}\}_i$ be an orthonormal basis for the orthogonal complement of $Q$. Then, at the end of the protocol, the joint state between $S$ and the measurement apparatus $A$ in the lab is of the form:

\be
\sum_{i} f_i(U)\ket{\psi(0)}_S\ket{\tilde{m}_i}_A,
\ee
\noindent where $f_i(U)$ are homogeneous polynomials of degree $4$ of the operators (\ref{partU}). No linear combination of these polynomials is central\footnote{Indeed, if $\sum_i c_if_i(U)$ were a central polynomial, then a projection on the state $\sum_i c^*_i |\tilde{m}_i>$ would have the effect of post-selecting system $S$ to $\psi(0)$. However, ${\cal W}_4$ is optimal when the prior on $U$ is the Haar measure, so we reach a contradiction.}. Under these circumstances, is there any way to return $S$ to its original state?

\begin{figure}
  \centering
  \includegraphics[width=6 cm]{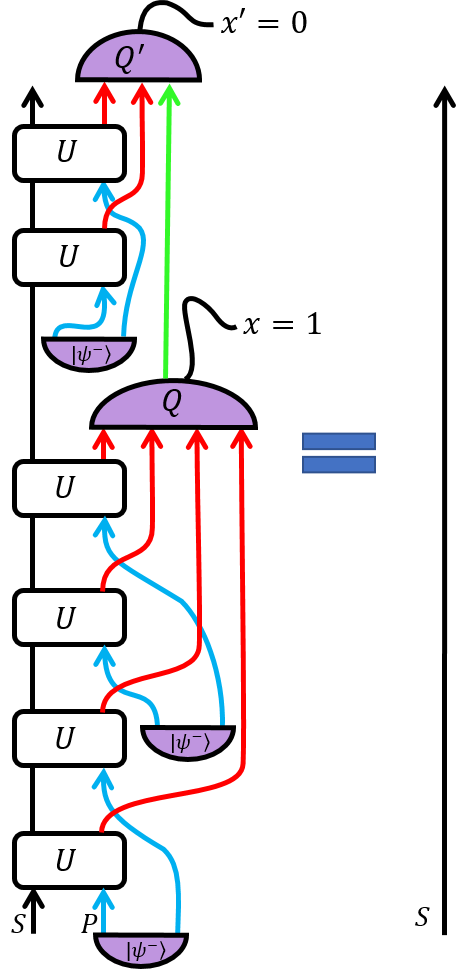}
  \caption{\textbf{Process diagram for the resetting of a failed resetting protocol.} The ressetting protocol depicted in Figure \ref{proto} has failed, i.e., $x=1$, but the last measurement was a non-demolition one, outputting the green system. We make two more probes in the singlet state interact with $S$ and then project them and the green output on a second space $Q'$. If the projection succeeds ($x'=0$), the system has been reset.}
  \label{undoing}
\end{figure}

Actually, there is. Suppose that we sent two more probes $P_1,P_2$ to $S$, say, in the singlet state and then we processed the state of the measurement apparatus together with these two probes, see Figure \ref{undoing}. If we project systems $A,P_1,P_2$ onto a pure state $\ket{m'}$, the final state of system $S$ will be

\be
\sum_{i}g_i(U)f_i(U)\ket{\psi(0)}_S,
\label{undoingF}
\ee
\noindent where $\{g_i(U)\}_i$ are homogeneous polynomials of degree $2$.

In principle it could be that, even though $\{f_i(U)\}_i$ are not central, there exist $\{g_i(U)\}_i$ such that $\sum_{i}g_i(U)f_i(U)$ is a central polynomial. Let $Q'$ be the space of tripartite measurement vectors in the two probes and the measurement apparatus inducing a central polynomial on $S$. If we implemented a projection on $Q'$ over $P_1,P_2,A$, conditioned on a successful outcome, we would undo, not only the unitary evolution under those two more time steps, but also the action of the previous resetting process, see Figure \ref{undoing}.

Using the tools developed in Appendices \ref{characterization}, \ref{heuristics}, we find that our theoretical speculations are sound (MATLAB code available in the Supplemental Material). The second row of Table \ref{rereset} shows the total probability of success of the resetting protocol after this second step. Note that, should this second measurement also fail ($x'=1$), we can make two more probes in the singlet state interact with the target and identify the controlled measurement that induces a central polynomial on $S$. Iterating this trick $m$ times, we end up at a resetting protocol whose duration is itself a dynamical variable, i.e., its exact value, that lies between $3T+4\delta$ and $(3+2m)T+(4+2m)\delta$, is determined during the course of the experiment.

Table \ref{rereset} presents the average probabilities of success of such concatenated resetting protocols for different values of $m$. The numbers between brackets reflect our precision in computing the average success probability over the Haar measure, which we replaced by a Montecarlo sampling of $100$ random unitaries $U$. Calculating the last row was a highly demanding computational task, as it involved characterizing subspaces of central polynomials spanned by thousands of vectors. The resulting truncated sequence of probabilities does not show any signs of saturation; for all we know, it may converge to $1$. The last value that we managed to estimate was $0.6585 \pm 0.0212$.


\begin{table}[tp]
\begin{tabular}{cc}
$m$ & average success probability\\
\hline
$0$ & $0.2053$ $(\pm 0.0142)$\\
$1$ & $0.2527$ $(\pm 0.0160)$\\
$2$ & $0.3663$ $(\pm 0.0196)$\\
$3$ & $0.4438$ $(\pm 0.0204)$\\
$4$ & $0.5300$ $(\pm 0.0215)$\\
$5$ & $0.5955$ $(\pm 0.0214)$\\
$6$ & $0.6585$ $(\pm 0.0212)$
\end{tabular}
\caption{Probability of success of concatenated resetting protocols. The numbers in brackets denote the statistical deviation of the estimated probabilities.}
\label{rereset}
\end{table}

\vspace{10pt}
\noindent\textbf{Conclusion}

We have proven that there exist probabilistic non-relativistic quantum protocols which allow one to bring an uncontrolled quantum system to a past state. These protocols work by making the uncontrolled system interact sequentially (in an unknown way) with a number of quantum probes, which are then processed in a controlled way. We showed that, should these protocols fail, one can then carry further protocols to undo the mess induced in the target and still drive it to its original state.

Our work raises several questions, the most important of which is how much more we can increase the average probability of success. It is a theoretical possibility that, as we consider protocols involving more and more probes, we manage to reset all possible unitary gates with probability arbitrarily close to $1$, except for a subset of measure zero. 

Another question concerns the duration of a resetting protocol. The protocol in Figure \ref{proto} requires more than $3T$ time units to send the target $T$ time units backwards in time. Is it feasible to reset the state of a quantum system $T$ time units by investing an arbitrarily short amount of time? Surely not, if we stick to the family of resetting protocols depicted in Figure \ref{reset}. We do believe, however, that an improvement in duration is possible, provided that we are allowed to exploit the which-path degree of freedom of the sent probes. It is also a topic for further research whether similar schemes can be devised to ``fast-forward'' the evolution of the target system, i.e., to achieve a time-warp experience with $\tau'> \tau$.

We conclude with this last reflection. We believe that the best thing one can do with a time-warping device is not to speculate about its theoretical features, but to \emph{turn it on}. Hence some experimental work towards an implementation of a resetting protocol would be welcome. In this regard, we feel that a future implementation of protocol ${\cal W}_4$ or some suboptimal variant, while challenging, is within reach of current quantum technologies.

\vspace{10pt}
\noindent\textbf{Acknowledgements}

\noindent We thank Fabio Costa for pointing to us that the Choi operator for a joint unitary evolution is a matrix product state: the realization that one can then apply cut-and-glue vectors to such an operator inspired the present work. We also acknowledge interesting discussions with Michal Sedlak, Zizhu Wang, Sukhwinder Singh and Susanne Berndl. This work was funded by the Austrian Science fund (FWF) stand-alone project P $30947$.

\begin{appendix}

\section{Resetting protocols for arbitrary $d_S$}
\label{existence}

The ideas behind protocol ${\cal W}_4$ can be generalized to show that there exist quantum resetting protocols for target systems of arbitrarily high dimension $d_S$ involving $O(d_S^3)$ qubit probes.

Indeed, suppose that we prepare $n$ probes in state $\ket{0}$, and make them interact sequentially via a unitary $V$ with a $d_S$-dimensional target. If we project the outgoing probes onto the state $\sum_{i_1,...,i_n}p^*_{i_1,...,i_n}\ket{i_1,...,i_n}$, the final wave function of the target will be $P(V_{0,0},V_{1,0})\ket{\psi(0)}$, where 

\be
V_{i,j}=(\id_S\otimes \bra{i})V(\id_S\otimes \ket{j})
\label{partV}
\ee

\noindent and $P(Y_0,Y_1)$ is the homogeneous matrix polynomial $P(Y_0,Y_1)=\sum_{i_1,...,i_n}p_{i_1,...,i_n}Y_{i_n}...Y_{i_1}$. To show that there exists a resetting protocol for system $S$, we will prove that there exists a polynomial $F$ such that, for any $2d_S\times 2d_S$ unitary $V$, $F(V_{0,0},V_{1,0})=f(V)\id_{d_S}$, with $f(V)$ vanishing just for a subset of unitaries of zero measure.

For any dimension $d_S$, there exists a homogeneous central polynomial $P(X)$ for dimension $d_S$ of degree $d_S^2$ involving $d_S+1$ matrix variables $X_1,...,X_{d_S+1}$ \cite{polyId}. Let $\bar{X}_1,...,\bar{X}_{d_S+1}$ be $d_S\times d_S$ matrices such that $P(\bar{X})\not=0$. In Appendix \ref{injectivity}, we show that there exists a unitary $U$ for $2d_S$-dimensional systems such that the products $\{U_{i_1,0}...U_{i_L,0}:i_1,...,i_L=0,1\}$ span $B(\C^{d_S})$, for $L=O(d_S)$. This implies, in particular, that there exist homogeneous matrix polynomials $\{f_i(Y_0,Y_1)\}_{i=1}^{d_S+1}$ of degree $L$ such that $f_i(U_{0,0},U_{1,0})=\bar{X}_i$ for $i=1,...,d_S+1$. It follows that the homogeneous central polynomial $F(Y_0,Y_1)\equiv P(f_1(Y_0,Y_1), f_2(Y_0,Y_1),...)$ of degree $O(d_S^3)$ is non-zero when evaluated on $(U_{0,0},U_{1,0})$. 

It just rests to show that $F(V_{0,0},V_{1,0})$ is also non-zero for generic unitaries $V$. For an arbitrary unitary $V\in B(\C^{2d_S})$, $F(V)\equiv F(V_{0,0},V_{1,0})=f(V)\id_{d_S}$, where $f(V)$ is an analytic function on the entries of $V$. If we parametrize $V$ via the generators of $\{R_j\}_{j=1}^{4d_S^2-1}$ of $SU(2d_S)$ as $V=e^{i\sum_{j}c_jR_j}$, then $f(V)=\tilde{f}(\vec{c})$, with $\tilde{f}$ analytic in $\R^{4d_S^2-1}$. Since $\tilde{f}$ is non-zero, by analyticity it can just vanish in a subset of $\R^{4d_S^2-1}$ of zero measure. Hence $f(V)$ is non-zero for generic $V$.

\section{Identifying injective unitaries}
\label{injectivity}
The purpose of this appendix is to prove that, for any $d_S\in\N$, there exists a unitary interaction $U\in B(\C^{2d_S})$ such that the products $\{U_{i_1,0}...U_{i_L,0}\}$ span $B(\C^{d_S})$, for $L=O(d_S)$.

Choose then two matrices $A_0,A_1\in B(\C^{d_S})$ such that, for some $L$, the products $\{A_{i_1}...A_{i_L}\}$ span the space $B(\C^{d_S})$. This can be done for $L=2d_S+1$, see Appendix A.1 in \cite{MPS}. Note that, by re-scaling and perturbing $A_0,A_1$, we can make sure that there exists $\rho>0$ with $\sum_i A^\dagger_i\rho A_i=\rho$. 

Let $\rho=R^2$, with $R>0$, and define the matrices $U_{i,0}\equiv R^{-1}A_i R$. It is trivial to see that the products $\{U_{i_1,0}...U_{i_L,0}\}$ span $B(\C^D)$ and that $\sum_{i=0,1}U^\dagger_{i,0}U_{i,0}=\id$.

Let us define the operator $U\in B(\C^{2d_S})$ on the subspace $\C^{d_S}\otimes \ket{0}$ via the relation $U\ket{\psi}\ket{0}=\sum_{i=0,1}U_{i,0}\ket{\psi}\ket{i}$. Since, within this subspace, $\bra{u}U^\dagger U\ket{v}=\braket{u}{v}$, there exists a unitary extension $\tilde{U}$ of $U$ to the whole space $\C^{d_S}\times\C^2$. By construction, $\tilde{U}_{i,0}=U_{i,0}$, and so the products $\{\tilde{U}_{i_1,0}...\tilde{U}_{i_L,0}\}$ span $B(\C^{d_S})$.

\section{Characterizing the set of resetting protocols}
\label{characterization}
The argument given in Appendix \ref{existence} cannot be used to devise practical quantum resetting protocols. In effect, taking the family of central polynomials proposed in \cite{polyId} and using the construction above invariably leads to protocols with a negligible average probability of success. What we need are methods which help us identify practical resetting schemes. This is the topic of the present Appendix.

Intuitively, everything amounts to making sure that, no matter how we interact with the probes, at the end of the protocol the wave function of system $S$ is multiplied by a central polynomial, or, at least, a matrix polynomial that is central when restricted to pieces of unitaries of the form (\ref{partU}). Building upon this idea, we provide a full characterization of the set of all quantum resetting protocols for fixed $n,d_S,d_P$. This characterization relies on the theory of quantum testers \cite{tester}, \cite{tester2} and on a variant of the method proposed in \cite{MPS} to compute the support of cut-and-glue operators for homogeneous matrix product states.

Take the quantum network depicted in Figure 2 in the main text, and fix the value of the unitary $U$. The action on the target system conditioned on a result $x=0$ is given by

\be
\tr_{IO}\left\{(\id_S\otimes (M_0)^T_{IO}) S_U\ket{\phi(0)}\bra{\phi(0)}S_U^\dagger\right\},
\ee
\noindent where $S_U$ is a rectangular operator of the form:

\be
S_U=\sum_{i_1,j_1,...,i_n,j_n}U_{j_n,i_n}...U_{j_1,i_1}\otimes\ket{i_1}_{I_1}\ket{j_1}_{O_1}...\ket{i_n}_{I_n}\ket{j_n}_{O_n}.
\ee
Here $O_k$ ($I_k$) denotes the Hilbert space of the $k^{th}$ output (input) qubit, and $O=\bigotimes_{k=1}^n O_k$ ($I=\bigotimes_{k=1}^n I_k$). $\{M_x\}_{x=0,1}$ are the Choi operators of our quantum tester \cite{tester}, \cite{tester2}. They are characterized by the conditions $M_x\geq 0, x=0,1$, $\sum_x M_x=\id_{O_n}\otimes \Gamma^{(n)}$, with $\tr_{I_{k}}(\Gamma^{(k)})=\id_{O_{k-1}}\otimes \Gamma^{(k-1)}$ \cite{tester}, \cite{tester2}.

Let $\sum_s\lambda_s\ket{c^s}\bra{c^s}$ be the spectral decomposition of $M^T_0\geq 0$. Since we want $\ket{\psi(0)}$ to remain the same, it follows that, for $s=1,...,2^{2n}$ and for all initial states $\ket{\psi(0)}$ and $U$'s, $(\id_S\otimes \bra{c^s})S_U$ must be proportional to the identity matrix. 

This shows why a generalization of resetting protocols to scenarios where $U$ represents a general trace-preserving completely positive map is impossible. Indeed, let $\{A^k\}_k$ be the Kraus operators of the map induced by $U$. After the action of the tester, the state $\rho_f$ of system $S$ would be a conic combination of states $\ket{\psi_{k_1,...,k_n}^s}$ of the form

\be
\ket{\psi_{k_1,...,k_n}^s}=\sum_{i_1,j_1,...,i_n,j_n}c^s_{i_1,j_1,...,i_n,j_n}A^{k_n}_{j_n,i_n}...A^{k_1}_{j_1,i_1}\ket{\psi(0)},
\ee
\noindent with $A^{k}_{j,i}=\id_S\otimes\bra{j}_P A^k\id_S\otimes\ket{i}_P$. Define now a quantum channel with Kraus operators $A_0=(\id-\proj{0})_S\otimes \id_P$, $A^{(i,j)}=\frac{1}{\sqrt{2}}\proj{0}\otimes\ket{j}\bra{i}$. Then, $\ket{\psi^{(i_1,j_1),...,(i_n,j_n)}(\bar{c})}=2^{-n/2}c^s_{i_1,j_1,...,i_n,j_n}\braket{0}{\psi(0)}\ket{0}$. If $c^s_{i_1,j_1,...,i_n,j_n}\not=0$, we thus have a term (in general) not proportional to $\ket{\psi(0)}$ in the decomposition of $\rho_f$. It is immediate to see that this also holds for small perturbations of the considered channel.

Coming back to unitary $U$s, the set of all vectors $\ket{c}$ such that, for all unitaries $U$, $(\id_S\otimes \bra{c^s})S_U$ is proptortional to the identity forms a vector subspace $\H^c\subset (\C^{2})^{\otimes 2n}$. In order to identify this subspace, we apply the same scheme used in \cite{MPS} to identify the related subspace of cut-and-glue vectors for matrix product states. Namely, we generate a sequence of states $(\ket{\xi^i})_i$ of the form $\ket{\xi}=(\bra{\varphi}_S\otimes \id_{IO}) S_{U}\ket{\varphi^\perp}_S$ by choosing random instances of $U, \varphi, \varphi^\perp$, with $\braket{\varphi}{\varphi^\perp}=0$. Applying the Gram-Schmidt method to the states $(\ket{\xi^i})_i$, we obtain an orthonormal basis for the subspace $\H^c_\perp$ spanned by them. The process is complete when the $k^{th}$ randomly generated vector lives in the span of the former $k-1$ vectors. This indeed indicates that $(\ket{\xi^i}_O)_{i=1}^{k-1}$ span $\H_\perp^c$, for, if they did not, the probability that a random vector in $\H_\perp^c$ belonged to their span would be zero.

$\H^c$ is the orthogonal complement of $\H^c_\perp$. Indeed, let $\ket{v}\perp \H^c$. Then, for fixed $U$, we have that, for all vectors $\ket{\psi_L}\perp\ket{\psi_R}$,

\be
(\bra{\psi_L}\otimes\bra{v})S_U\ket{\psi_R}=0.
\ee
\noindent It follows that $(\id_{S}\otimes\bra{v})S_U=f(U) \id_S$, where $f(U)$ is a scalar. This must hold for all $U$, so $\ket{v}\in H^c$. The opposite implication is immediate.

Now, suppose that we have a prior distribution $\rho(U)dU$ of unitaries $U$, and we want to maximize the average probability of success $\int \rho(U)p(x=0,|U,\pi)dU$ over all quantum resetting protocols $\pi$ involving $n$ probes. By all the above, this reduces to the following problem:

\begin{align}
&\max\tr(M^T_0X(\rho))\nonumber\\
\mbox{s.t. }&\mbox{supp}(M^T_0)\subset \H^c,M_0,M_1\geq 0\nonumber\\
&M_0+M_1=\id_{A_n^{out}}\otimes \Gamma^{(n)}, \nonumber\\
&\tr_{I_{k}}(\Gamma^{(k)})=\id_{O_{k-1}}\otimes \Gamma^{(k-1)},
\label{SDP_char}
\end{align}
\noindent with $X(\rho)=\int dU \rho(U) W(U)W(U)^\dagger$, $W(U)=(\bra{0}_S\otimes\id_{IO})S_U\ket{0}$.

\noindent This is a semidefinite program \cite{sdp}, and, as such, we can solve it in time polynomial on $2^n$. Taking $n=4$, $d_S=2$ and using the numerical packages MOSEK \cite{mosek} and YALMIP \cite{yalmip}, we find that the maximum average success probability for the prior $\rho(U)dU=dU$ coincides with the one achieved by protocol ${\cal W}_4$ (up to a precision of $10^{-6}$). This suggests that ${\cal W}_4$ is an optimal or extremal resetting protocol.

\section{Devising new resetting protocols}
\label{heuristics}
The implementation of the SDP (\ref{SDP_char}) for $n>4$ turned out to be computationally prohibitive, mainly due to lack of computer memory. Hence we had to rely on heuristics to devise new resetting schemes. We settled on \emph{prepare-and-measure protocols}, where we input an $n$-qubit state into the network and then measure the $n$ output qubits in some appropriate basis. Note that the optimal protocol ${\cal W}_4$ falls in this category, so perhaps this restriction is not that limiting after all.

In the following, we present two heuristics to find extremal measure-and-prepare resetting protocols for high $n$. Given a guess $\ket{\varphi}$ on the input state, heuristic $\# 1$ returns the projection operator $\Pi$ with maximum support such that the pair $(\varphi,\Pi)$ constitutes a sound quantum resetting protocol. Conversely, given a guess on the final projection operator $\Pi$, heuristic $\#2$ returns the state space $\H_{\Pi}$, such that, for any $\varphi\in \H_{\Pi}$, the pair $(\varphi,\Pi)$ is a valid resetting protocol. Given a prior $\rho(U)dU$, identifying the state $\varphi^\star\in \H_{\Pi}$ that maximizes the average success probability reduces to an eigenvalue problem.

For extreme prepare-and-measure strategies, $M_0=\proj{\varphi}_I\otimes \Pi^T_O$, where $\ket{\varphi}$ is the input state and $\Pi$ is the projector describing the measurement of the output qubits. To devise a resetting protocol, we need to make sure that the support of $M^T_0$ is in $\H^c$. 

Suppose that we have a guess on the $n$-qubit input state $\ket{\varphi}$. In analogy with the characterization of $H^c$, the maximum support of $\Pi$ corresponds to the orthogonal complement of the subspace $\H^c_{\varphi}$ spanned by vectors of the form

\be
(\bra{\psi_L}\otimes\bra{\varphi^*}_I\otimes\id_O)S_U\ket{\psi_R}, 
\label{vectos}
\ee
\noindent with $\ket{\psi_L}\perp\ket{\psi_R}$. The latter can be characterized using the randomizing algorithm described before. Note that there may exist vectors $\ket{m}$ with $\ket{\varphi^*}\otimes\ket{m}\in \H^c$ such that, for all unitaries $U$, $\id_S\otimes\bra{\varphi^*}_I\bra{m}_OS_U$ is the $d_S\times d_S$ null matrix. The corresponding measurements do not contribute to the final average probability of success, so such vectors can be eliminated. This can be achieved by considering the space $\H_{\varphi}$ spanned by vectors of the form (\ref{vectos}), with $\ket{\psi_L}=\ket{\psi_R}$. The measurement space we are interested in is thus $\H_\varphi\cap \H^c_{\varphi}$.

Similarly, given a guess on the support of $\Pi$, we can find the subspace $H_\Pi$ of input states such that $\varphi\in H_\Pi$ implies $\mbox{supp}(\proj{\varphi^*}\otimes \Pi)\subset \H^c$.

We first used heuristic $\# 1$ to explore the scenario $d_S=d_P=2,n=8$. We identified the protocols ${\cal W}_8$, $\tilde{{\cal W}}_8$, with input states $\ket{\varphi}=\bigotimes_{k=1}^4\ket{\psi^{-}}_{2k-1,2k}$ and $\ket{\tilde{\varphi}}=\ket{\psi^{-}}_{13}\ket{\psi^{-}}_{24}\ket{\psi^{-}}_{57}\ket{\psi^{-}}_{68}$, respectively, and corresponding projective measurements $\Pi,\tilde{\Pi}$ of ranks $39$ and $78$. The MATLAB code employed is in the Supplemental Material.

Let us test the performance of these new protocols in a specific physical scenario. Let system $S$ evolve freely via the Hamiltonian $H_0=\lambda Z$ for time $T$, and interact with system $P$ through the term $-(X+Z)_S\otimes Y_P$. We assume that probes access and leave the system very quickly, and that they are quantum memories, i.e., $H_P=0$. The full interaction between the probe and the target is thus modeled by the Hamiltonian $H_I=-(X+Z)\otimes Y+ \lambda Z\otimes \id$, and the total unitary interaction is $U(\lambda,T)=e^{-iH_{INT}\delta}e^{-iH_0T}$. We take $\lambda$ to be distributed uniformly in the interval $[-1,1]$, and $\delta=0.5$.

Note that the above assumptions are just required to compute the average probability of success. Even if all of them turned out to be false, any resetting protocol we will consider next would be sound. That is, conditioned on $x=0$, it would reset $S$ to $\ket{\psi(0)}$, as long as the total unitary interaction $U_{SP}$ between the target and the probe is the same during the $n$ steps of the protocol.

\begin{figure}
  \centering
  \includegraphics[width=8 cm]{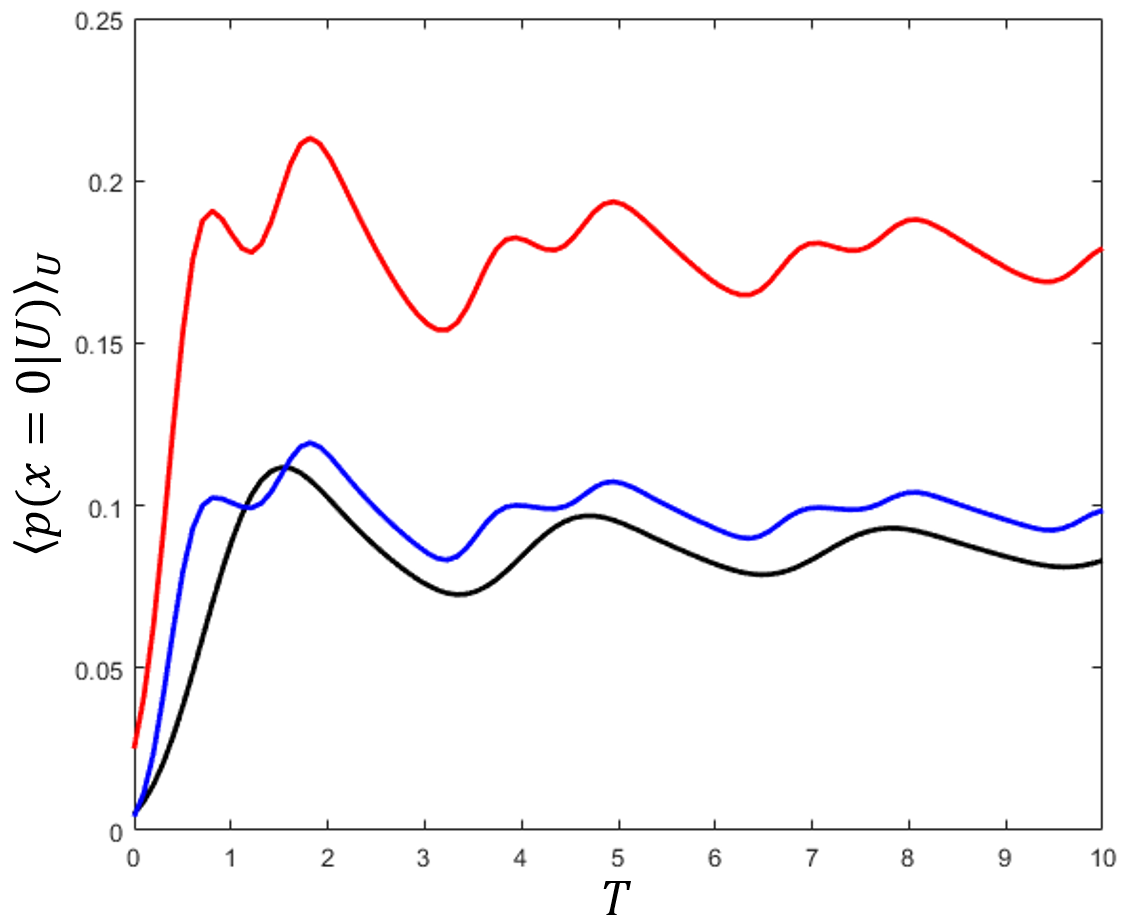}
  \caption{\textbf{Probability of success as a function of the resetting time $T$ for and interaction generated by the Hamiltonian $H_{INT}$, with $\delta=0.5$.} The black, blue and red lines represent, respectively, the average probabilities of success achieved by the optimal $n=4$ protocol, ${\cal W}_8$ and $\tilde{{\cal W}}_8$.}
  \label{plot1}
\end{figure}

Figure \ref{plot1} compares the results of strategies ${\cal W}_8,\tilde{{\cal W}}_8$ with the best available protocol for $n=4$ probes, computed via SDP. The curves obtained are peculiar in that, at the beginning, the probability of success increases the further in the past we want to revert the target. We also see that ${\cal W}_8,\tilde{{\cal W}}_8$ supersede all protocols involving four probes, at the price of doubling the duration of the whole resetting procedure.

\begin{figure}
  \centering
  \includegraphics[width=8 cm]{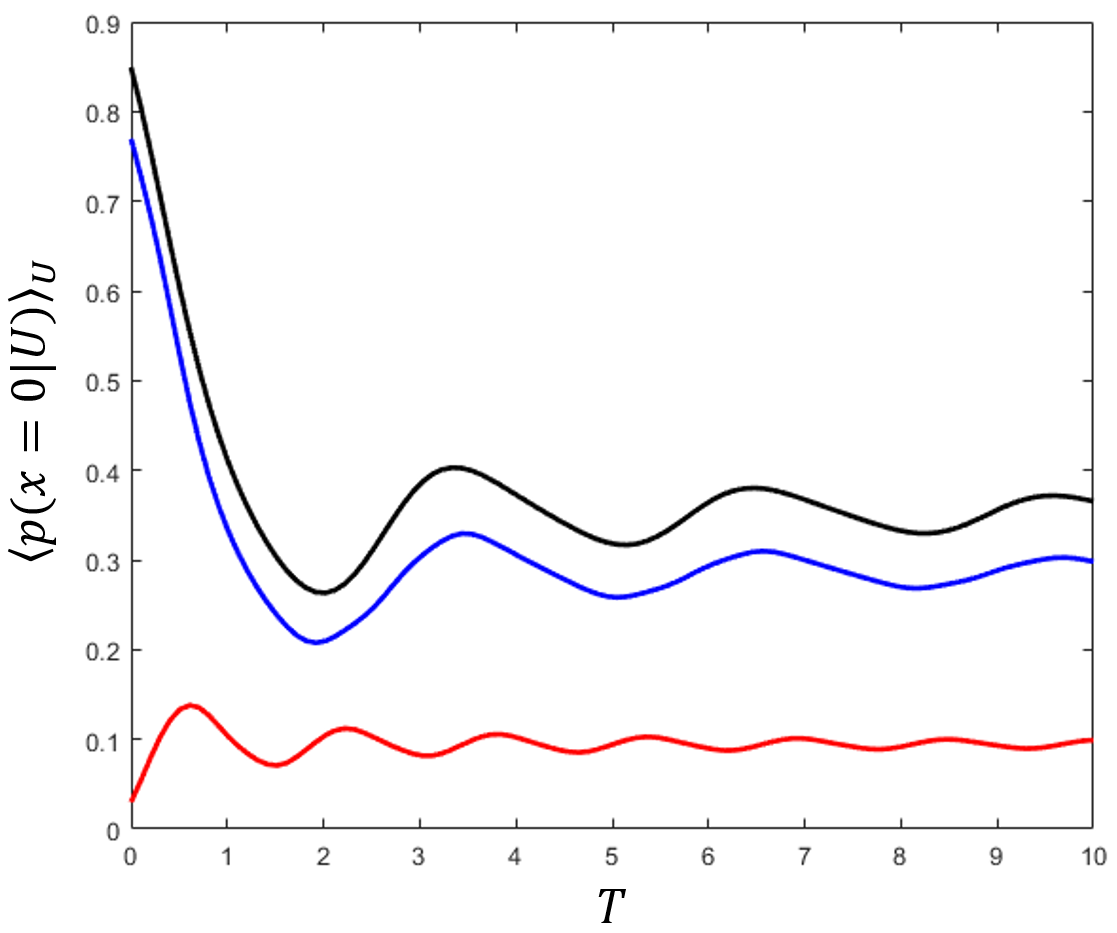}
  \caption{\textbf{Probability of success as a function of the resetting time $T$ for and interaction generated by the Hamiltonian $H'_{INT}$, with $\delta=1$.} The black, blue and red lines represent, respectively, the average probabilities of success achieved by the optimal $n=4$ protocol, ${\cal W}_8$ and $\tilde{{\cal W}}_8$.}
  \label{plot2}
\end{figure}

We must not conclude, though, that ${\cal W}_8,\tilde{{\cal W}}_8$ are superior to, say, ${\cal W}_4$. Let us keep the same free Hamiltonian $H_0(\lambda)$, again with $\lambda$ distributed uniformly in $[-1,1]$, but change the interaction Hamiltonian $H_{INT}$ to $H'_{INT}=-0.7464X\otimes X+1.4885Y\otimes Y-3.1014Z\otimes Z + \lambda Z\otimes \id$ (i.e., $U(\lambda,T)=e^{-iH'_{INT}\delta}e^{-iH_0T}$). For $\delta=1$, we obtain the curves depicted in Figure \ref{plot2}.

\noindent Now $n=4$ protocols supersede both ${\cal W}_8$ and $\tilde{{\cal W}}_8$, and the latter protocol achieves the worst results. The take-home message is that no single protocol outperforms all the others in all situations. The decision to use one strategy or another will depend on our prior knowledge on the target and its interaction with the probes.

To conclude, we tried to identify some resetting protocols for $d_S=3$. This turned out to be very challenging, because there exist no central polynomials for dimension $3$ with degree smaller than $8$ \cite{polyId}. Moreover, for $n=8$ there seems not to be any resetting protocol involving qubit probes. For $n=9$, the only protocol we found, ${\cal W}_9$, involves projecting the output qubits on the symmetric space of $n$ qubits, and inputting a vector determined by the heuristic $\# 2$. Our prior on $U$ was a discrete distribution given by $100$ $6\times 6$ unitaries sampled from the Haar measure. Putting all together, we obtain the discouragingly small value $\langle p(x=0|U,{\cal W}_9)\rangle_U=0.0035(\pm 0.0004)$.

\end{appendix}

\bibliography{timeMachine}

\end{document}